\documentclass[aps,prl,reprint,twocolumn,superscriptaddress]{revtex4-2}
\usepackage{graphicx}
\usepackage{amsmath}
\usepackage{hyperref}
\usepackage{color}

\begin{document}

\title{First Evidence for an Unambiguous Triangle Singularity from
$\psi(2S) \to p\bar{p}\eta$}

\author{Qi Huang}
\email{06289@njnu.edu.cn}
\affiliation{Department of Physics and Technology, Nanjing Normal University, Nanjing 210023, China}

\author{Yi-Jia Zeng}
\email{zengyj@ihep.ac.cn}
\affiliation{University of Chinese Academy of Sciences, Beijing 100049, China}
\affiliation{Institute of High Energy Physics, Chinese Academy of Sciences, P.O. Box 918(1), Beijing 100049, China}

\author{Xiao-Rui Lyu}
\email{xiaorui@ucas.ac.cn, corresponding author}
\affiliation{University of Chinese Academy of Sciences, Beijing 100049, China}

\author{Rong-Gang Ping}
\email{pingrg@ihep.ac.cn, corresponding author}
\affiliation{University of Chinese Academy of Sciences, Beijing 100049, China}
\affiliation{Institute of High Energy Physics, Chinese Academy of Sciences, P.O. Box 918(1), Beijing 100049, China}

\author{Jia-Jun Wu}
\email{wujiajun@ucas.ac.cn, corresponding author}
\affiliation{University of Chinese Academy of Sciences, Beijing 100049, China}

\date{\today}

\begin{abstract}
Triangle singularities, predicted by Landau in 1959, are purely kinematic enhancements arising from hadronic rescattering loops.
Despite their proposed role in various anomalous decay processes and exotic hadron candidates, a direct experimental confirmation has remained elusive for more than six decades.
We analyze the $p\eta/\bar{p}\eta$ invariant mass
spectrum in $\psi(2S) \to p\bar{p}\eta$ measured by the BESIII Collaboration.
The data exhibit a clear cusp-like structure around 1.564~GeV in the $N(1535)$ region, in precise agreement with the kinematic position predicted for the triangle singularity.
Including the triangle singularity loop in the fit substantially improves the description of the data,
with $\chi^2/\mathrm{d.o.f.}$ decreasing from 1.22 to 0.90, corresponding to a significance of $\sim 3.8\sigma$ for the triangle singularity contribution.
The precise alignment of the observed excess with the predicted kinematic position provides the first compelling evidence for the triangle singularity effect.
\end{abstract}

\maketitle

\paragraph{Introduction}

In hadron spectroscopy, distinguishing genuine resonances from kinematic effects remains a major challenge.
The triangle singularity (TS) is a prominent example of such a kinematic effect.
First predicted by Landau in 1959~\cite{Landau1960}, TSs are purely kinematic singularities of the scattering amplitude.
Unlike dynamical resonances, whose pole positions depend on interaction details, the location of a TS is fixed by the masses of the three internal particles in the loop and those of the initial- and final-state particles.
Under the Coleman--Norton theorem~\cite{Coleman1965,Shen2020}, a TS can manifest as a
logarithmic singularity on the physical boundary.
It produces a peak or a dip that closely mimics a resonance. This masquerading behavior poses a serious challenge to establishing an unambiguous hadron spectrum~\cite{Guo2020}.

For decades, experimental information on TSs was scarce.
The first observed peak attributed to a TS was the large isospin-violating signal in
$J/\psi \to \gamma\eta(1405/1475) \to \gamma\pi^0 f_0(980) \to \gamma + 3\pi$~\cite{Li2011,Ablikim2012}.
A triangle loop formed by $\eta(1405) \to K^*\bar{K}K \to
\pi^0 f_0(980)$ was introduced to argue that the TS could induce significant isospin
violation~\cite{Wu2012,Aceti2012,Wu2013,Achasov2015,Du2019}.
This work brought the TS back
into focus.
Subsequently, TS contributions have been proposed for many processes, including those with near-threshold exotic candidates such as $Z_c$~\cite{Wang2013a,Wang2013b,
Szczepaniak2015,Pilloni2017,Liu2013,Yu2024,Chen2024,Zhang2025},
$P_c$~\cite{Guo2016,Liu2016,Guo2015,Bayar2016}, and $X(2900)$~\cite{Liu2020,Burns2021}
(see Ref.~\cite{Guo2020} for a review).
Yet, a direct and unambiguous experimental confirmation of a TS remains elusive.
The main difficulty is that the TS position often lies too close to a resonance threshold.
It can also be obscured by the large widths of intermediate states.
Both effects hinder a clean separation of kinematic and dynamic contributions~\cite{Liu2016b}.

A telling example is the $a_1(1420)$ signal observed by the COMPASS Collaboration in the $f_0(980)\pi$ final state.
This signal emerges on top of the broad $a_1(1420)$ resonance.
It was initially interpreted as a new exotic meson.
Later, the narrow peak was argued to originate from a TS in the $a_1(1420) \to K^*\bar{K} \to f_0(980)\pi$ rescattering
chain~\cite{Alexeev2021,Mikhasenko2015}.
The TS model describes the data with fewer parameters and a slightly better fit quality than the resonance hypothesis~\cite{Alexeev2021}.
This shows how subtle the distinction between a TS-induced peak and a genuine resonance can be.
As Aitchison remarked in his commentary on the COMPASS result, ``this is not unambiguous evidence for the observation of a triangle singularity, but the paper shows
pretty convincingly that it is sufficient to explain the data, and that a new resonance is not required''~\cite{CERN2021}.

To detect a clean TS, a systematic analysis~\cite{Liu2016b,Huang2021} has identified three
criteria:
(i)~the TS position must be well separated from threshold enhancements,
(ii)~the intermediate particles must have narrow widths, and
(iii)~the TS contribution must be quantitatively calculable.
Applying these criteria, five years ago, Ref.~\cite{Huang2021} proposed that the $\psi(2S) \to p\bar{p}\eta$ process
satisfies all three conditions.
In this process, the TS arises from the rescattering loop
$\psi(2S)\to J/\psi \eta \to p\bar{N}^*+N^* \bar{p} \to p\bar{p}\eta$ (with $N^*$ denoting the $N(1535)$ resonance).
Its kinematic peak is predicted to appear at $m_{p\eta} = 1.56387$~GeV, far above the $p\eta$ threshold and within the dominant $N(1535)$ signal region.
The initial proposal assumed a vanishing relative phase between the tree-level and loop diagrams.
Whether the TS would be quantitatively visible was therefore left open.

Recently, the BESIII Collaboration reported a high-precision measurement of $\psi(2S)\ \to p\bar{p}\eta$~\cite{Ablikim2025}, providing an ideal dataset to test this prediction.
In this Letter, we analyze the $p\eta/\bar{p}\eta$ invariant mass spectrum from Ref.~\cite{Ablikim2025}.
We show that the data exhibit a clear cusp-like structure around 1.564~GeV, which aligns precisely with the predicted TS position.
We also find that including the TS loop diagram in the fit significantly improves the description of the data, with $\chi^2/\mathrm{d.o.f.}$ decreasing from 1.22 to 0.90.
This corresponds to a significance of $\sim 3.8\sigma$ for the TS contribution.
These results therefore constitute the first unambiguous evidence for the triangle singularity effect.

\paragraph{Formalism}

\begin{figure}[t]
\centering
\includegraphics[width=\columnwidth]{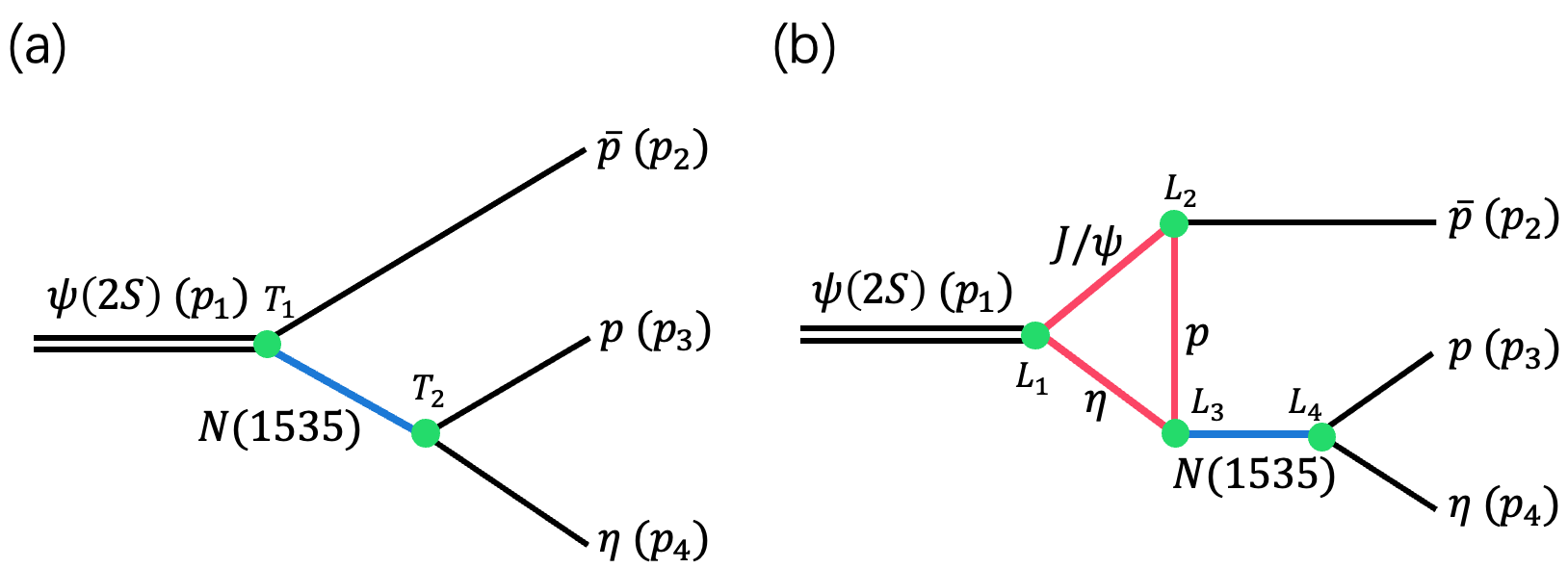}
\caption{The Feynman diagrams for $\psi(2S) \to p\bar{p}\eta$.
(a)~Tree diagram dominated by the intermediate $N(1535)$.
(b)~Loop diagram where the triangle singularity occurs.
We use the experimental BW description of $N(1535)$ as input~\cite{Ablikim2025}.}
\label{fig:diagrams}
\end{figure}

The process $\psi(2S) \to p\bar{p}\eta$ receives contributions from both the direct decay
(tree diagram) and the hadronic rescattering loop, as shown in Fig.~\ref{fig:diagrams}. The
$p\eta$ final-state interaction is dominated by the $N(1535)$ resonance, described by a
Breit-Wigner (BW) function~\cite{Ablikim2025,Hunt2019}:
\begin{align}
\mathrm{BW}(m, m_0, \Gamma_0) &= \frac{1}{m^2 - m_0^2 + i m_0 \Gamma(m, m_0, \Gamma_0)},
\label{eq:bw}\\
\Gamma(m, m_0, \Gamma_0) &= \Gamma_0 \sum_i r_i \frac{\rho_i(m)}{\rho_i(m_0)}.
\label{eq:width}
\end{align}
Here, $m$ is the invariant mass, $m_0$ and $\Gamma_0$ are the nominal mass and width of
$N(1535)$, and $\Gamma(m, m_0, \Gamma_0)$ includes the energy-dependent partial widths.
The branching fractions $r_i$ of the $i$-th decay channel are $r_1 = r_2 = 0.43$ for
$N(1535) \to N\eta/N\pi$, and $r_3 = 0.14$ for $N(1535) \to N\rho$. The phase-space factor
$\rho_i$ is parameterized as $\rho_{1/2} = q_{1/2}/m$ for the stable $N\eta$ or $N\pi$
channel, where $q_i$ is the momentum of the two particles in their center-of-mass frame.
For the $\rho$ meson channel, $\rho_3 = \int \frac{q_3}{m} \mathrm{BW}(m, m_\rho,
\Gamma_\rho) \, dm$ owing to the finite width of the $\rho$ meson.

The tree-level and loop amplitudes for the diagrams in Fig.~\ref{fig:diagrams} are
\begin{align}
\mathcal{M}_{\text{Tree}} &= T_2 T_1 \,
\mathrm{BW}(m_{34}, m_{N(1535)}, \Gamma_{N(1535)}), \label{eq:mtree}\\
\mathcal{M}_{\text{Loop}} &= i \int \frac{d^4q}{(2\pi)^4}
\frac{L_3 F(p_3 + p_4 - q, m_\eta, \Lambda_\eta)}{(p_3 + p_4 - q)^2 - m_\eta^2 + i m_\eta \Gamma_\eta}
\nonumber\\
&\quad \times \frac{L_1 F(p_2 + q, m_{J/\psi}, \Lambda_{J/\psi})}{(p_2 + q)^2 - m_{J/\psi}^2 + i m_{J/\psi} \Gamma_{J/\psi}}
\nonumber\\
&\quad \times \frac{L_2 F(q, m_p, \Lambda_p)}{q^2 - m_p^2 + i m_p \Gamma_p}
\, L_4 \, \nonumber\\
&\quad \times \mathrm{BW}(m_{34}, m_{N(1535)}, \Gamma_{N(1535)}). \label{eq:mloop}
\end{align}
Here, the vertex interactions $T_i$ and $L_i$ are extracted from the following Lagrangians:
\begin{align}
\mathcal{L}_{\psi(2S)J/\psi\eta} &=
g_{\psi(2S)J/\psi\eta} \, \varepsilon^{\mu\nu\alpha\beta} \,
\partial_\mu \psi(2S)_\nu \partial_\alpha \psi_\beta \eta, \label{eq:lag1}\\
\mathcal{L}_{\psi p\bar{p}} &= -g_{\psi p\bar{p}} \, \bar{p}\gamma^\mu \psi_\mu p,
\label{eq:lag2}\\
\mathcal{L}_{\eta p N(1535)} &= -g_{\eta p N(1535)} \, \bar{p} N(1535) \eta + \text{h.c.},
\label{eq:lag3}
\end{align}
where the relevant parameters can be found in Ref.~\cite{Huang2021}. The coupling
$g_{\eta p N(1535)} = 2.59 \pm 0.62$ is determined from the branching ratio
$\mathcal{B}(N(1535) \to p\eta) = (30-55)\%$.

The $T_1$ vertex for $\psi(2S) \to \bar{p}N(1535)$ is expressed, following the BESIII
Collaboration~\cite{Ablikim2025}, as
\begin{equation}
T_1 = i g_1 \epsilon^\mu_{\psi'} \bar{u}(p_{\bar{p}})
\bigl(\gamma_5 \sigma_{\mu\nu} p^\nu_{\psi'} f_{94}
+ \gamma_5 \gamma_\mu f_{93}\bigr) v(p_{N(1535)}), \label{eq:t1}
\end{equation}
where $g_1$ is a free parameter, and the BESIII Collaboration provides
$f_{93} = -11.34 - 5.24i$ and $f_{94} = 0.66 + 0.42i$~\cite{Ablikim2025}.
Since $f_{94}$ is much smaller and contributes only at higher order, being suppressed by
the baryon masses~\cite{Zou2003}, we neglect it in the following calculation. The magnitude
of $g_1$ is estimated by
$(g_{\psi(2S)\bar{p}N(1535)} \times g_{N(1535)p\eta}) / (g_{N(1535)p\eta} |f_{93}|)$,
where $(g_{\psi(2S)\bar{p}N(1535)} \times g_{N(1535)p\eta}) = (1.38 \pm 0.07) \times
10^{-3}$ is calculated from the experimental branching ratio
$\mathcal{B}(\psi(2S) \to \bar{p}N(1535) + \mathrm{c.c.} \to \bar{p}p\eta)
= (4.7 \pm 0.5) \times 10^{-5}$.

In the loop amplitude of Eq.~\eqref{eq:mloop}, we introduce a form factor
$F(q, m, \Lambda) = \Lambda^4 / [(q^2 - m^2)^2 + \Lambda^4]$ to describe the structure
effects of the interaction vertices and the off-shell effects of the internal particles.
This form factor also regularizes the ultraviolet divergence. The cutoffs are parameterized
as $\Lambda_{J/\psi,\eta,p} = m_{J/\psi,\eta,p} + \alpha \Lambda_{\text{QCD}}$, with
$\alpha = 1$~\cite{Huang2021} and $\Lambda_{\text{QCD}} = 0.22$~GeV.
At the TS point, all internal particles are on-shell, which keeps the form factor close to unity.
Consequently, the cusp structure is largely insensitive to the specific choice of $\alpha$.

The total amplitude is a coherent superposition of the tree and loop contributions:
\begin{equation}
\mathcal{M}_{\text{Total}} = C \bigl( \mathcal{M}_{\text{Tree}}
+ r e^{i\phi} \mathcal{M}_{\text{Loop}} \bigr), \label{eq:mtotal}
\end{equation}
where $C$ normalizes the overall rate, and $r$ and $\phi$ control the
relative magnitude and phase of the two amplitudes. The differential decay width is
obtained by squaring the total amplitude and integrating over phase space.

\paragraph{Results and Discussion}

\begin{figure}[t]
\centering
\includegraphics[width=\columnwidth]{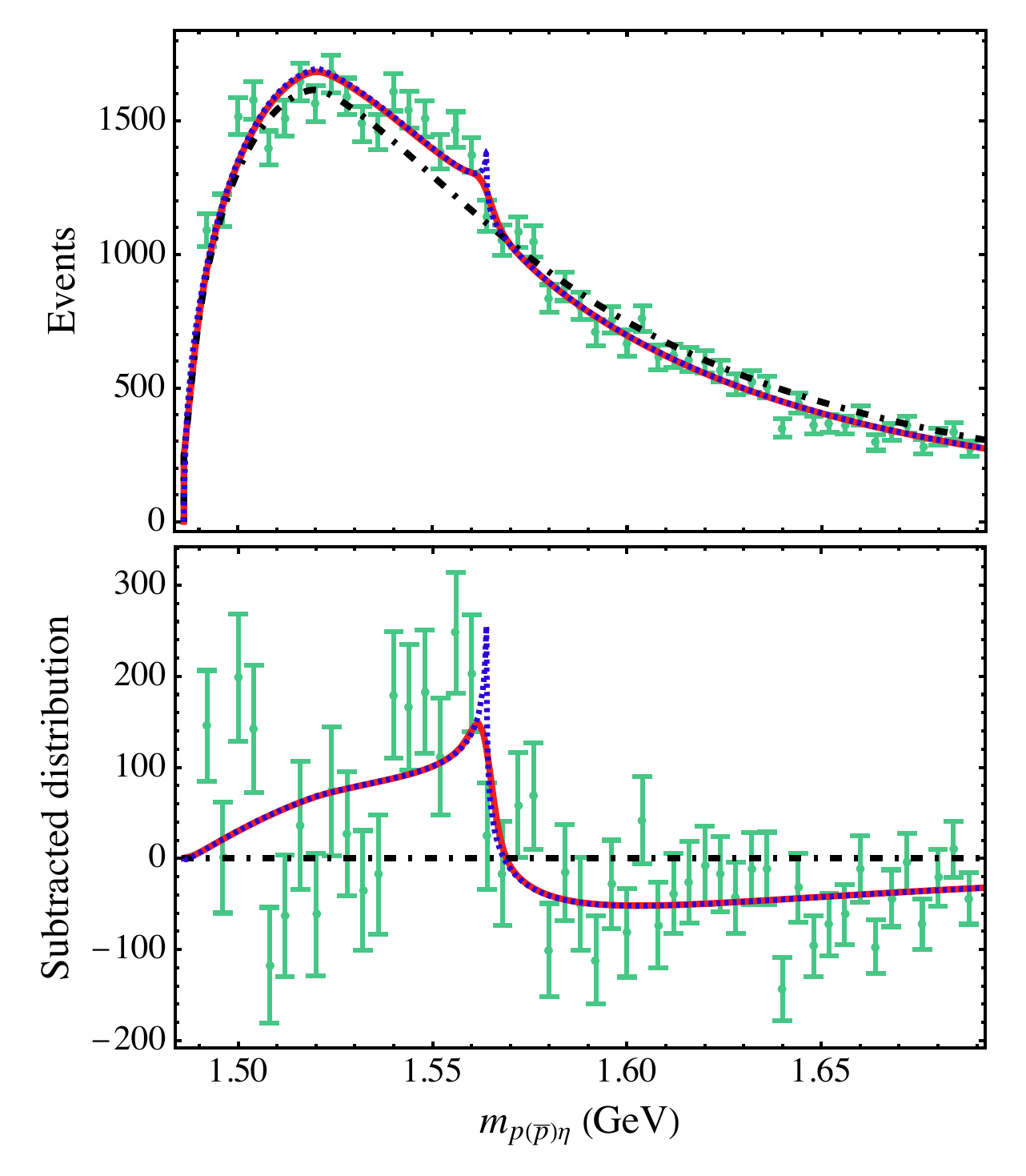}
\caption{
Upper panel: Fit to the $p\eta$/$\bar{p}\eta$ invariant mass spectrum from BESIII~\cite{Ablikim2025}.
The black dash-dotted line shows the pure $N(1535)$ BW fit ($\chi^2/\mathrm{d.o.f.}=1.22$).
The red solid line includes the TS loop contribution with convolution of the detector resolution  ($\chi^2/\mathrm{d.o.f.}=0.90$).
The purple dotted line shows the same without convolution, revealing the sharp of the TS cusp.
Lower panel: Data and fits after subtracting the fitted $N(1535)$ BW distribution.
}
\label{fig:fit}
\end{figure}

In the upper panel of Fig.~\ref{fig:fit}, we show the sum of measured invariant mass spectrum of
$p\eta$ and $\bar{p}\eta$ in the range $1.48-1.69$~GeV as green circles, where each event is only counted for one time.
The background has been subtracted and the efficiency correction has been applied.
Since the contribution from $J/\psi$ is removed by requiring $|m_{p\bar{p}} - 3.097| > 0.050$~GeV~\cite{Ablikim2025}, the $N(1535)$ resonance dominates the experimental data.

Furthermore, in practical the experimental detector must have a finite resolution.
To account for the resolution effect, we convolve our fit result with a Gaussian function,
\begin{eqnarray}
    \gamma(m_{p\eta}) \to \frac{1}{\sqrt{2\pi}\sigma}\int_{m_{p\eta}-3\sigma}^{m_{p\eta}+3\sigma}\gamma(m) e^{-\frac{(m_{p\eta}-m)^2}{2\sigma^2}} dm.
\end{eqnarray}
Here, $\gamma(m_{p\eta}) \equiv d\Gamma(m_{p\eta})/dm_{p\eta}$ is the differential decay width, and $\sigma = 2$~MeV is the detector resolution, as provided by the BESIII Collaboration~\cite{Ablikim2025}.

We first fit the $p\eta/\bar{p}\eta$ invariant mass spectrum using only the tree-level amplitude.
The fitted parameters are listed in Table~\ref{tab:params}. %
The resulting $\chi^2/\mathrm{d.o.f.}$ is 1.22, indicating a reasonable but imperfect description of the data.
As shown by the black dot-dashed curve in the upper panel of Fig.~\ref{fig:fit}, the data points around $1.54-1.56$~GeV lie systematically above the BW curve.
A pronounced cusp-like excess appears near 1.564~GeV.
This coincides precisely with the predicted TS
position, $m_{p\eta} = 1.56387$~GeV~\cite{Huang2021}.

We then include the loop diagram in the fit, with $r$ and $\phi$ as free parameters.
The fit quality improves substantially, as the purple dotted line shows, with $\chi^2/\mathrm{d.o.f.}$ decreasing to 0.90.
For $N_{\text{data}} = 50$ bins in the fit range, this corresponds to $\Delta\chi^2 = 17.0$ for $\Delta\mathrm{d.o.f.} = 2$.
Under the asymptotic $\chi^2$ approximation, this corresponds to a significance of $\sim3.8\sigma$ for the TS contribution,
based on the $p$-value of a $\chi^2$ improvement test~\cite{Navas2024}.
The phase $\phi = -0.25$, as listed in Table~\ref{tab:params}, is close to
zero, as anticipated in Ref.~\cite{Huang2021}, indicating constructive interference.

A comparison with the data is shown in Fig.~\ref{fig:fit}.
The red solid line, which includes the TS contribution together with the convolution, provides a significantly better description of the $\sim1.55$~GeV region than the black dash-dotted line.
Moreover, removing the energy-resolution convolution (purple dotted line) renders the sharp TS peak structure highly visible, in excellent agreement with the theoretical expectation of Ref.~\cite{Huang2021}.
To highlight the improvement around the TS region, we subtract the fitted $N(1535)$ BW distribution from both the experimental data and the fitted results.
As shown in the lower panel of Fig.~\ref{fig:fit},
the TS structure captured by the full (tree~+~loop) fit is clearly visible. When above about 1.57~GeV, the curve falls below zero, consistent with the data tendency.
Thus, the constructive interference between the tree and loop amplitudes enhances the spectrum around 1.54--1.56 GeV and drives it below the BW baseline above 1.57 GeV, consistent with the dip observed in the experimental data as given in the lower panel of Fig.~\ref{fig:fit}.

\begin{table}[t]
\centering
\caption{Fit parameters for the $p\eta$ invariant mass spectrum. The symbol ``$\times$''
indicates that the parameter is not used in the corresponding fit.}
\label{tab:params}
\begin{tabular}{lcc}
\hline
\hline
Parameter & $N(1535)$ & $N(1535) + \text{Loop}$ \\
\hline
$C$      & $8.90 \pm 0.04$ & $8.10 \pm 0.12$ \\
$r$      & $\times$        & $1.92 \pm 0.28$ \\
$\phi$   & $\times$        & $-0.25 \pm 0.12$ \\
\hline
$\chi^2/\mathrm{d.o.f.}$ & 1.22 & 0.90 \\
\hline
improvement & \multicolumn{2}{c}{$3.8\sigma$}\\
\hline
\hline
\end{tabular}
\end{table}

The deviation of $r$ from unity deserves comment.
In our framework, the absolute magnitude of the single-loop diagram is essentially fixed by known coupling constants and
kinematics~\cite{Huang2021}.
The fitted value $r = 1.92$ may suggest that multi-loop
corrections further enhance the TS effect. This is consistent with the observation that triangle singularities can be significantly amplified by multi-loop mechanisms~\cite{Nakamura2024}.
It indicates that the TS signal in $\psi(2S) \to p\bar{p}\eta$ may be even more robust than a naive single-loop estimate would suggest.

Taken together, our results provide first unambiguous evidence for the triangle singularity effect in the $\psi(2S) \to p\bar{p}\eta$ process. A full toy-Monte-Carlo validation and a systematic uncertainty evaluation with higher-statistics
data will provide a more definitive assessment.

\paragraph{Summary}

We have presented the first unambiguous evidence for the triangle singularity effect, predicted by Landau more than six decades ago~\cite{Landau1960}.
We notice a cusp-like structure in the BESIII data at 1.564~GeV in the $p\eta$ invariant mass spectrum of $\psi(2S) \to p\bar{p}\eta$.
The position aligns precisely with the kinematic prediction~\cite{Huang2021}.
After including the TS effect, the $\sim 3.8\sigma$ preference from the fit constitutes a
direct validation of the hadronic loop mechanism in QCD.
While a higher-statistics dataset will be needed for a definitive confirmation, the present results provide the first unambiguous identification of a TS signature, free from resonance-threshold ambiguities.

Our results demonstrate that purely kinematic singularities can manifest as distinct, observable structures in hadronic decays.
This is a crucial step toward disentangling kinematic effects from genuine resonances in hadron spectroscopy.
With more data accumulating at BESIII~\cite{yu2026sixteen}, this evidence is expected to become definitive
confirmation of the triangle singularity effect.

\paragraph{Acknowledgments}
We thank the BESIII Collaboration for providing the data used in this work.
This work is supported by the National Key Research and Development Program of China under
Contract No.~2025YFA1613900,
and by the National Natural Science Foundation of China under Grants No.~12221005 and No.~12305087,
and by the Chinese Academy of Sciences under Grant
No.~YSBR-101.

\end{document}